\documentstyle[aps,pre,epsf]{revtex}
\begin{document}
%
%
%
%
\draft
\twocolumn[\hsize\textwidth\columnwidth\hsize\csname @twocolumnfalse\endcsname
\title{\Large \bf Damage Spreading in the Ising Model}
\author{Haye Hinrichsen and Eytan Domany}
\address{Department of Physics of Complex Systems,
	 Weizmann Institute, Rehovot 76100, Israel}
\date{cond-mat/xxyyzzz, revised version, March 11, 1997}
\maketitle
\begin{abstract}
We present two new results regarding damage spreading in 
ferromagnetic Ising models. First, we show that a damage 
spreading transition can occur in an Ising chain that evolves 
in contact with a thermal reservoir. Damage heals at low
temperature and spreads at high $T$. The dynamic  rules for 
the system's evolution for which such a transition is observed
are as legitimate as the conventional rules (Glauber, Metropolis, 
heat bath). Our second result is that such transitions are
not always in the directed percolation universality class.
\end{abstract}]
\pacs{{\bf PACS numbers:} 05.50.q, 05.70.Ln, 64.60.Ak, 64.60.Ht \\
      {\bf Key words:} \hspace{6mm} Ising model, damage spreading}
%
%
%
%
%
%
\section{Introduction}
A system is said to exhibit damage spreading  (DS) if the ``distance'' between  
two of its replicas, that evolve under the same thermal noise but from
slightly different initial conditions, increases with time. 
Even though DS was first introduced in the 
context of biologically motivated dynamical systems
\cite{Kauffman}, it has evolved into 
an important tool in physics. It is used in equilibrium \cite{z}
for measuring accurately dynamic exponents and also out of equilibrium,
to study the influence of initial conditions on the temporal evolution 
of various systems. In particular, one hoped that DS
could be used to identify ``phases'' of {\it chaotic} 
behavior in systems with no intrinsic dynamics, 
such as Ising ferromagnets~\cite{Creutz,Stanley} and
spin-glasses\cite{DerWeis}. Such hopes were dampened when it
was realized that different algorithmic implementations of the 
same physical system's dynamics
(such as Glauber versus heat bath or Metropolis Monte Carlo)
can have different DS properties \cite{Mariz,JandArc}.
This implies that DS is not an intrinsic
property of a system\cite{GrJSP79}, 
since two equally legitimate algorithms yield
contradictory results. This problem was addressed  
recently in \cite{PreviousPaper}, where we  realized that one {\it can}
define ``phases'' on the basis of their DS properties
in an algorithm-independent manner. 
To do this one must, however, consider simultaneously the 
{\it entire set} $\cal A$
of possible algorithms (dynamic procedures) that are consistent with the 
physics  of the model studied (such as detailed balance, interaction
range and symmetries). Every system 
must belong to one of three possible DS phases, depending
on whether damage spreads for all, none or a part of the members of the set
$\cal A$.

Once we have been led to consider a large family of algorithms, it 
was natural to revisit an old question, such as the 
possibility for DS in the one-dimensional (1-$d$) Ising 
ferromagnet. In this case all conventional dynamic procedures agree 
that damage does not spread. We show here that  once the family of dynamic
procedures is extended in the spirit explained above, {\it a DS transition is 
possible in the 1-$d$ Ising model.} Having found such a DS transition, it 
is again natural to investigate to which universality class it belongs. So far
this issue could be addressed only for the 2-$d$ case; since it 
is much easier to obtain high-quality numerical data in 1-$d$, 
we were able to test carefully a conjecture of Grassberger 
\cite{GrJSP79}, according to which 
the generic universality class of damage spreading transitions 
is directed percolation (DP). This indeed is correct, but we discovered that 
if the dynamics that is being used has certain symmetries, 
{\it the DS transition is not in the DP class}. Interestingly this is the 
case for Glauber dynamics of the $H=0$ Ising model, 
for which the DS transition is non-DP.

We start by reviewing briefly~\cite{Mariz,JandArc} 
the conventional algorithms - Glauber, heat bath (HB) and Metropolis - 
and show that they form a 
particular subset of some 
general set of legitimate rules $\cal A$. All members of $\cal A$
satisfy detailed balance
with respect to the same Hamiltonian; hence all these rules
generate the same equilibrium 
ensemble as the conventional algorithms and
are equally legitimate to 
mimic the temporal evolution of an Ising system in contact 
with a thermal reservoir.
Next, we introduce two ``new'' dynamic rules, which 
constitute just another subset 
of $\cal A$, and show that for these two rules
a DS transition  {\it does} occur in the
1-$d$ Ising model. Moreover, as we show in the example of the second rule,
an additional $Z_2$ symmetry of the DS order parameter leads to a transition
that is not in the DP universality class. 

\section{Previous work, with conventional algorithms}
Denote the site  which is being updated by $i$
and the set of its neighbors by $j$. The energy at time $t$ is given by
\begin{equation}
\frac{{\cal H}}{k_BT} = -\sum_ih_i(t)\sigma_i(t),
\hspace{5mm}
h_i(t) = \sum_j K_{ij} \sigma_j(t), \nonumber
\end{equation}
where $K_{ij}=J/k_BT$ and
$\sigma_i(t)=\pm 1$.
Define a transition probability $p_i(t)$:
\begin{equation}
p_i(t)= \frac{e^{h_i(t)}}{e^{h_i(t)}+e^{-h_i(t)}}.
\label{eq:pi}
\end{equation}
The update rules of HB, Glauber and Metropolis dynamics
are expressed in terms of random numbers 
$z=z_i(t)$, selected with equal probability from the interval $[0,1]$. 
The rule for {\it standard HB} is
\begin{equation}
\label{StandardHB}
\sigma_i(t+1)={\rm sign}[p_i(t)-z].
\end{equation}
A different dynamic process is obtained by
generating at each site {\it two independent} random numbers, 
$z_+$ and $z_-$, and using the first if $\sigma_i(t)=+1$ and 
the second when $\sigma_i(t)=-1$. The rules of this {\it uncorrelated HB}
dynamics may be written as
\begin{equation}
\label{UncorrHB}
\sigma_i(t+1)=
\left\{ \begin{array}{ll}
{\rm sign}[p_i(t)-z_+] & \mbox{if $ \sigma_i(t)=+1$} \\
{\rm sign}[p_i(t)-z_-] & \mbox{if $ \sigma_i(t)=-1$}
\end{array} \right. .
\end{equation}
{\it Glauber} dynamics uses only one random number per site:
\begin{equation}
\label{Glauber}
\sigma_i(t+1)=
\left\{ \begin{array}{ll}
+{\rm sign}[p_i(t)-z] & \mbox{if $ \sigma_i(t)=+1$} \\
-{\rm sign}[1-p_i(t)-z] & \mbox{if $ \sigma_i(t)=-1$}
\end{array} \right. .
\end{equation}
This rule can be expressed in the form of (\ref{UncorrHB}) but with the
two random numbers completely {\em anticorrelated}, i.e., $z_++z_-=1$.

Finally the rules for {\it Metropolis} dynamics read
\begin{equation}
\label{Metropolis}
\sigma_i(t+1)=
\left\{ \begin{array}{ll}
+{\rm sign}[p^+_i(t)-z] & \mbox{if $ \sigma_i(t)=+1$} \\
-{\rm sign}[p^-_i(t)-z] & \mbox{if $ \sigma_i(t)=-1$}
\end{array} \right. ,
\end{equation}
where $p^\pm_i(t)=\min(1,e^{\mp 2 h_i(t)} )$.

It is easy to show that given $\sigma_{i-1}(t),\sigma_i(t),\sigma_{i+1}(t)$,
the probability to get $\sigma_i(t+1)= +1$ 
is the same for standard HB, uncorrelated HB, and Glauber 
dynamics\footnote{For Metropolis dynamics the transition probabilities
are different}.
Hence, by observing the temporal evolution of a {\it single}
Ising system, one cannot tell by which of these 
methods was its trajectory in configuration 
space generated.

\begin{table}
\small
\begin{tabular}{|c||c|c|c|c|c|}
correlation & Glauber & usual & uncorr. & dynamics & dynamics \\
function & dyn.  & HB & HB & of eq.(11) & of eq.(18)\\
\hline  \hline
$\langle r_{---}\,r_{--+} \rangle $&
	$1-\kappa$ & $1-\kappa$ & $1-\kappa$ & $\kappa-1$ & $\lambda(1-\kappa)$
	\\
$\langle r_{---}\,r_{-+-} \rangle $&
	$2 \kappa -1$ & $1$ & $\kappa^2$ & $1$ & $2 \kappa -1$ \\
$\langle r_{---}\,r_{-++} \rangle $&
	$\kappa-1$ & $1-\kappa$ & $0$ & $\kappa-1$ & $\lambda(\kappa-1)$ \\
$\langle r_{---}\,r_{+-+} \rangle $&
	$1-2\kappa$ & $1-2\kappa$ & $1-2\kappa$ & $1-2\kappa$ & $1-2\kappa$\\
$\langle r_{---}\,r_{+++} \rangle $&
	$-1$ & $1-2\kappa$ & $-\kappa^2$ & $1-2\kappa$ & $-1$ \\
$\langle r_{--+}\,r_{-+-} \rangle $&
	$\kappa-1$ & $1-\kappa$ & $0$ & $\kappa-1$ & $\lambda(\kappa-1)$ \\
$\langle r_{--+}\,r_{-++} \rangle $&
	$-1$ & $1$ & $0$ & $1$ & $-1$ \\
$\langle r_{--+}\,r_{+--} \rangle $&
	$1$ & $1$ & $1$ & $1$ & $1$\\
$\langle r_{--+}\,r_{+-+} \rangle $&
	$1-\kappa$ & $1-\kappa$ & $1-\kappa$ & $\kappa-1$ & $\lambda(1-\kappa)$
	\\
$\langle r_{--+}\,r_{++-} \rangle $&
	$-1$ & $1$ & $0$ & $1$ & $-1$ \\
$\langle r_{-+-}\,r_{+-+} \rangle $&
	$-1$ & $1-2\kappa$ & $-\kappa^2$ & $1-2\kappa$ & $-1$ \\
\end{tabular}
\caption{Two-point correlations in the one-dimensional Ising model
for various dynamic rules.
We used the notation $\kappa = \tanh \frac{2J}{k_BT}$.}
\label{TableGlauberHB}
\end{table}

The difference between these dynamics may become 
evident only when we observe the evolution of two replicas, i.e.,
study damage spreading!
Indeed Stanley et al \cite{Stanley} and also Mariz et al \cite{Mariz}
found, using Glauber dynamics, that damage spreads
for the 2-$d$ Ising model for $T>T_c$; similarly
for Metropolis dynamics \cite{Mariz}. More recently Grassberger
\cite{GrassJPA} claimed that the DS transition occurs
slightly below $T_c$ for Glauber which 
was also observed in the corresponding
mean field theory \cite{Vojta}. 
On the other hand damage does not
spread at any temperature with standard HB 
for neither the 2-$d$ \cite{Mariz} nor the
3-$d$ Ising models \cite{DerWeis}. 
The 3-$d$ model did exhibit DS for
$T>T^*$ with $T^*<T_c$ when Metropolis \cite{Costa} and Glauber
\cite{GrassJPA,LaCaer} dynamics were used.
In the 1-$d$ Ising model with HB, Glauber or Metropolis dynamics
no damage spreading has been observed.

\section{General class of dynamic procedures for the Ising model}
The dynamic rules considered here for the $1$-$d$ Ising model consist
of local updates, where a random variable $r=\pm1$ is 
assigned to the spin $\sigma_i$:
\begin{equation}
\sigma_i(t+1) := r_{\sigma_{i-1}(t),\sigma_{i}(t),\sigma_{i+1}(t)}\,.
\end{equation}
This random variable is generated in some probabilistic procedure
using one or several random numbers. Like in the conventional algorithms
discussed above, we allow 
the random variable to depend only on the values taken at time $t$ by
the updated spin itself and the spins with which it interacts 
(i.e., its nearest 
neighbors). The set of all one-point 
functions $\langle r_{\sigma_{i-1},\sigma_{i},\sigma_{i+1}} \rangle$ 
determines the transfer matrix of a {\it single} system.
Here $\langle \ldots \rangle$ denotes the average over many 
independent realizations of random numbers. 
The simultaneous evolution (and, hence, DS) of {\it two replicas} 
$\{ \sigma \}$ and $\{ \sigma' \}$ is, however, governed by a joint
transfer matrix of the two systems which, in turn, 
is completely determined by the two-point functions 
$\langle r_{\sigma_{i-1},\sigma_{i},\sigma_{i+1}}
         r_{\sigma'_{i-1},\sigma'_{i},\sigma'_{i+1}} \rangle$. 
In general, $n$-point functions determine the joint 
transfer matrix of $n$ replicas. An important requirement is that
all correlation functions have to
be invariant under the symmetries of the model~\cite{PreviousPaper}. 
For a homogeneous Ising chain
in zero field these symmetries are invariance under reflection
\begin{eqnarray}
\langle r_{\sigma_{i-1},\sigma_i,\sigma_{i+1}} \rangle &=&
\langle r_{\sigma_{i+1},\sigma_i,\sigma_{i-1}} \rangle \,,\\
\langle r_{\sigma_{i-1},\sigma_i,\sigma_{i+1}} 
r^\prime_{\sigma^\prime_{i-1},\sigma^\prime_i,
\sigma^\prime_{i+1}} \rangle &=&
\langle r_{\sigma_{i+1},\sigma_i,\sigma_{i-1}} 
r^\prime_{\sigma^\prime_{i+1},\sigma^\prime_i,
\sigma^\prime_{i-1}} \rangle \nonumber \,,
\end{eqnarray}
and global inversion of all spins ($Z_2$ symmetry):
\begin{eqnarray}
&&\langle r_{\sigma_{i-1},\sigma_i,\sigma_{i+1}} \rangle =
-\langle r_{-\sigma_{i-1},-\sigma_i,-\sigma_{i+1}} 
\rangle \nonumber \,, \\
&&\langle r_{\sigma_{i-1},\sigma_i,\sigma_{i+1}} 
r^\prime_{\sigma^\prime_{i-1},\sigma^\prime_i,
\sigma^\prime_{i+1}} \rangle =\\
&& \hspace{1cm} \langle r_{-\sigma_{i-1},-\sigma_i,
-\sigma_{i+1}} r^\prime_{-\sigma^\prime_{i-1},
-\sigma^\prime_i,-\sigma^\prime_{i+1}} \rangle \nonumber \,.
\end{eqnarray}
For both HB and for Glauber dynamics the one-point functions are given by
\begin{equation}
\label{OnePointCorrelations}
\langle r_{\sigma_{i-1},\sigma_{i},\sigma_{i+1}} \rangle =
2 p_i-1.
\end{equation}
The corresponding transfer matrices for single systems
are, hence, identical.
On the other hand the two-point functions for HB and Glauber
dynamics are different so that damage
evolves differently (see Table I). 
Still, damage does not spread in $1$-$d$ for any
of these algorithms at any temperature.

\section{Dynamic rule for which damage does spread in 1-d}
Consider the following dynamics for the $1$-$d$ Ising model:
\begin{equation}
\label{NeighborDynamics}
r_{\sigma_{i-1},\sigma_i,\sigma_{i+1}} =
\left\{
\begin{array}{ll}
+{\rm sign}(p_i-z) & \mbox{if} \ \sigma_{i-1} = \sigma_{i+1} \\
-{\rm sign}(1-p_i-z) & \mbox{if} \ \sigma_{i-1} \neq \sigma_{i+1}
\end{array}
\right.
\end{equation}
As can be checked easily, this dynamical rule yields the same one-point
correlations as in eq. (\ref{OnePointCorrelations}). Therefore,
the evolution of a single replica using this rule  cannot be 
distinguished from that of Glauber or HB dynamics. However, the
two-point correlations (and therewith damage spreading properties) 
are different (see Table I).

\begin{figure}
\epsfxsize=85mm
\epsffile[60 600 540 770]{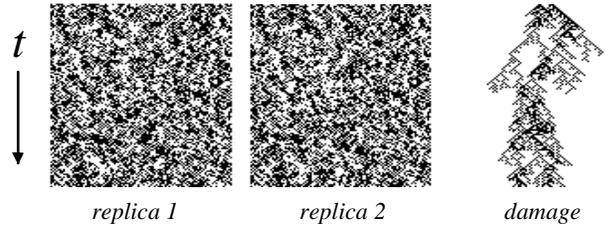}
\caption{ Temporal evolution of damage in the 1-$d$ Ising model of size $200$
with the dynamics of eq.~(11) near the DS transition $J/k_BT^*$=0.2305.  
Each configuration is represented by a row of pixels and time goes downwards.
The two replicas are started from identical initial conditions. At an
early time, a damage of 5 sites is inserted in the center. }
\label{FigureOne}
\end{figure}

Unlike Glauber and HB, this  dynamics does exhibit
a damage spreading transition in $1$-$d$. This can be seen as follows.
At $T=\infty$ eq. (\ref{NeighborDynamics}) reduces to
\begin{equation}
r_{\sigma_{i-1},\sigma_i,\sigma_{i+1}} = 
\sigma_{i-1}\sigma_{i+1}\,{\rm sign}(\frac12-z)\,,
\end{equation}
which implies
that the local damage 
$\Delta_i(t)=1-\delta_{\sigma_i(t),\sigma^\prime_i(t)}$
evolves deterministically:
\begin{equation}
\label{InfiniteTemperature}
\Delta_i(t+1) =
\left\{
\begin{array}{ll}
0 & \mbox{if} \ \Delta_{i-1}(t) = \Delta_{i+1}(t) \\
1 & \mbox{if} \ \Delta_{i-1}(t) \neq \Delta_{i+1}(t)
\end{array}
\right.
\end{equation}
Since this is exactly the update rule of a 
Domany-Kinzel model \cite{DomanyKinzel}
in the active phase (with $p_1=1$ and $p_2=0$), we conclude that 
for $T=\infty$ damage spreads. On the other hand, for $T=0$
eq. (\ref{NeighborDynamics}) reduces to
\begin{equation}
\label{ZeroTemperature}
r_{\sigma_{i-1},\sigma_i,\sigma_{i+1}} =
\left\{
\begin{array}{ll}
\sigma_{i-1} & \mbox{if} \ \sigma_{i-1}=\sigma_{i+1} \\
{\rm sign}(z-\frac12) & \mbox{if} \ \sigma_{i-1} \neq \sigma_{i+1}
\end{array}
\right. \, .
\end{equation}
In this case damage evolves probabilistically and cannot be viewed as an
independent process. One can, however, show that the expectation value 
to get damage at site $i$, averaged over many realizations of random numbers
satisfies the inequality
$\langle \Delta_i(t+1)\rangle \leq \frac12
 \langle \Delta_{i-1}(t)+\Delta_{i+1}(t) \rangle$, that is
$\langle \Delta(t+1) \rangle \leq \langle \Delta(t) \rangle$. This
means that for $T=0$ damage does not spread. In fact, simulating the
spreading process one observes a DS transition at finite
temperature. 
A typical temporal evolution near the transition is 
shown in Fig. \ref{FigureOne}. 

In order to determine the critical exponents
that characterize the DS transition, we perform dynamic
Monte-Carlo simulations \cite{DynamicMC}. Two replicas
are started from

%
%
\noindent
%
%
%
%
%
\begin{figure}
\epsfxsize=90mm
\epsffile[70 430 520 770]{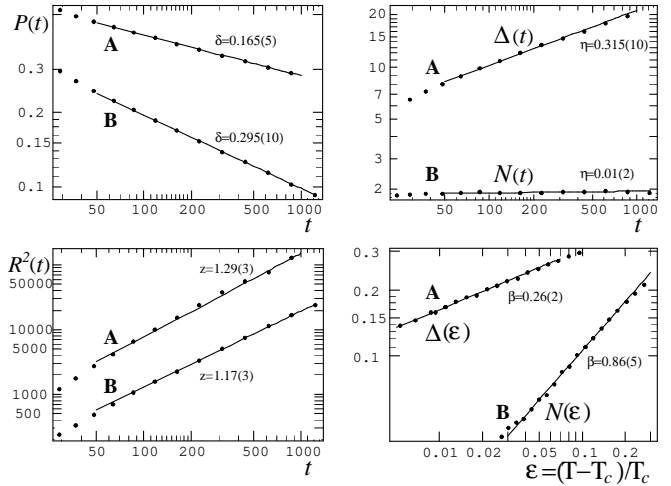}
\caption{Numerical results for the 1-$d$ Ising model
with {\bf A:}~the dynamics of eq.~(11) and {\bf B:} 
the dynamics of eq.~(18). The measured quantities 
are explained in the text.}
\label{FigureTwo}
\end{figure}
\noindent
identical random initial conditions, where one
damaged site is inserted at the center. Both replicas then evolve
according to the dynamic rules of the system using the same set of
random numbers. In order to minimize finite-size
effects, we simulate a large system of $5000$ sites with periodic boundary
conditions. For various temperatures we perform $10^6$ independent runs
up to $1500$ time steps. However, in many runs damage heals very soon
so that the run can be stopped earlier. As usual in this type of
simulations, we measure the survival probability $P(t)$, the number
of damaged sites $\Delta(t)$, and the mean-square spreading of damage from
the center $R^2(t)$ averaged over the active runs. At the DS transition,
these quantities are expected to scale algebraically in the large time
limit:
\begin{equation}
\label{ScalingQuantities}
P(t) \sim t^{-\delta} \,, \hspace{10mm}
\Delta(t) \sim t^\eta \,, \hspace{10mm}
R^2(t) \sim t^z\,.
\end{equation}
The critical exponents $\delta,\eta,z$
are related to the density exponent $\beta$
and the scaling exponents $\nu_\perp$, $\nu_{||}$ by
$\delta = \beta/\nu_{||}$, $z = 2 \nu_\perp /\nu_{||}$
and obey the hyperscaling relation $4\delta+2\eta = dz$.
At criticality, the quantities (\ref{ScalingQuantities}) 
show straight lines in double logarithmic plots. 
Off criticality, these lines are curved. Using
this criterion we estimate the critical temperature for the
DS transition by $J/k_BT^*=0.2305(5)$. The exponents $\delta$, $\eta$, 
and $z$ are measured at criticality while the density exponent $\beta$ 
is determined off criticality by measuring the stationary Hamming distance 
$\Delta(T) \sim (T-T^*)^\beta$ in the spreading phase. 
The results of our simulations are
shown in Fig. \ref{FigureTwo}. From the slopes in the double
logarithmic plots we obtain the estimates
$\delta=0.165(5)$, $\eta=0.315(10)$, $z=1.29(3)$, 
and $\beta=0.26(2)$ which are in fair agreement with the 
known~\cite{JensenPRL96} exponents for directed percolation
$\delta=0.15947(3)$, $\eta=0.31368(4)$, $z=1.26523(4)$, and $\beta=0.27649(4)$.
We therefore conclude that in agreement with 
Grassberger's conjecture \cite{GrJSP79}, the DS transition belongs to
the DP universality class.
This is very plausible; as far as the damage variable is concerned 
there is a {\it single absorbing state} (of no damage at all) and the 
transition is from a phase in which the system ends up in this state 
to one in which it does not, just as is the case for DP.

\section{Damage spreading transition with non-DP exponents}
Different critical properties are expected 
\cite{GrassbergerAB,Menyhard,MonDim,BARW,TwoAbsStates,Cardy}
for rules with two
distinct absorbing states (of the damage variables!) related by symmetry.
It is important to note that the $Z_2$ symmetry of the Ising system
does not suffice - inverting all spins in {\it both} replicas does not
change the damage variable 
(the Hamming distance between the two configurations). Therefore, 
we are looking for dynamic rules which (a) have two types of absorbing states; 
one with no damage and the
other with full damage. Furthermore, (b) the two  play completely
symmetric roles. One can see that both (a) and (b) 
hold for rules that satisfy the condition 
\begin{equation}
\label{Z2Symmetry}
r_{\sigma_{i-1},\sigma_i,\sigma_{i+1}} = 
-r_{-\sigma_{i-1},-\sigma_i,-\sigma_{i+1}}\,.
\end{equation}
The immediate consequence of this condition is that if a configuration
$\{ \sigma (t) \}$ evolves in one time step into $\{  \sigma (t+1) \}$,
then the spin-reversed configuration $\{ -\sigma (t) \}$ will evolve into
precisely   $\{  -\sigma (t+1) \}$. Imagine now simultaneous evolution of
two replicas with initial states $\{ \sigma \}$ and $\{ \sigma' \}$, 
giving rise
to a damage field $\{ \Delta \}$. Reversal of the initial state on {\it one} 
of the replicas will give sign-reversed spin states on this replica and hence
the damage field   $\{ -\Delta \}$ will evolve. 
Thus, for rules
that satisfy condition (16), the {\it damage variable} has an $Z_2$ symmetry. 
A particular consequence of this symmetry is that if two initial states
are the exact sign-reversed of one another, this will persist at all subsequent
times. Therefore, inasmuch as  $\Delta =0$ (no damage) is an absorbing state,
so is the situation of full damage, $\Delta = 1$. For systems with such
$Z_2$ symmetry we expect the DS transition (if it exists) to exhibit non-DP
behavior.  

%
%
%
\begin{figure}
\epsfxsize=85mm
\epsffile[60 580 530 770]{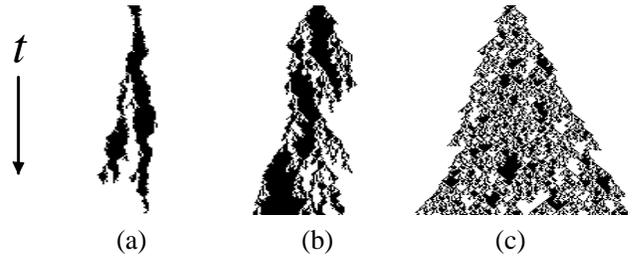}
\caption{$Z_2$-symmetric damage spreading transition. 
Two replicas with 200 sites are started from 
identical random initial conditions. At an early time 5 damaged sites
are introduced in the center. For fixed 
temperature $J/k_BT=0.25$ a typical temporal evolution of damage is shown for
(a) Glauber dynamics $\lambda=1$, (b) near the transition $\lambda^*=0.82$
and (c) in the spreading regime $\lambda=0$. Because of the symmetry,
islands of damaged sites can heal only at the edges.}
\label{FigureThree}
\end{figure}

\noindent

It is quite remarkable to note that Glauber dynamics satisfies 
eq. (\ref{Z2Symmetry})! The $Z_2$-symmetry of damage 
in the 1-$d$ Glauber model is illustrated in Fig. \ref{FigureThree}a.
One can see that compact islands of damaged sites are formed
because damage does not heal spontaneously inside such islands
but only at the edges. However, as mentioned earlier, there is no DS
transition in the 1-$d$ Glauber model.

Consider now a different dynamic rule:
\begin{equation}
\label{PCSpreadDynamics}
r_{\sigma_{i-1},\sigma_i,\sigma_{i+1}} =
\left\{
\begin{array}{ll}
+{\rm sign}(p_i-z) & \mbox{if} \ \sigma_{i-1}\sigma_{i}\sigma_{i+1} = 1 \\
-{\rm sign}(1-p_i-z) & \mbox{if} \ \sigma_{i-1}\sigma_{i}\sigma_{i+1} =-1
\end{array}
\right.
\end{equation}
For this rule, which also
satisfies eq. (\ref{Z2Symmetry}), 
we observe
in simulations that damage always spreads
(see Fig. \ref{FigureThree}c).
In order to generate an $Z_2$-symmetric
DS transition in $1$-$d$, we use a rule that 
interpolates between this and Glauber. 
This can be done by introducing a second parameter 
$0 \leq \lambda \leq 1$
and `switching' between Glauber dynamics and rule (\ref{PCSpreadDynamics})
as follows: in each update an additional random number $\tilde{z}$ 
is generated. If $\tilde{z} \geq \lambda$, rule (\ref{PCSpreadDynamics}) 
is applied, otherwise Glauber dynamics is used. 
This mixed dynamics can be expressed as
\begin{equation}
\label{MixDynamics}
r_{\sigma_{i-1},\sigma_i,\sigma_{i+1}} =
\left\{
\begin{array}{ll}
+{\rm sign}(p_i-z) & \mbox{if} \ y = 1 \\
-{\rm sign}(1-p_i-z) & \mbox{if} \ y =-1 \\
\end{array}
\right. \,,
\end{equation}
where $y=\frac12 \sigma_i[(1+\sigma_{i-1}\sigma_{i+1})+
(1-\sigma_{i-1}\sigma_{i+1}) {\rm sign}(\lambda-\tilde{z})]$.
Again this rule leads to the one-point correlations of 
eq. ~(\ref{OnePointCorrelations}), i.e., the temporal evolution of a single
replica is the same as in Glauber and HB dynamics. However,
varying $\lambda$ (at fixed  T) we find a critical value $\lambda^*$
where a DS transition occurs.
A typical temporal evolution of damage near the transition is
shown in Fig.~\ref{FigureThree}b.

Since `damage' and `no damage' play a symmetric role,
the Hamming distance $\Delta$ (the density of damaged sites) cannot
be used as an order parameter. Instead one has to use the
density of {\em kinks} $N$ (domain walls) between damaged and healed
domains. By definition, the number of kinks is conserved modulo two
which establishes a parity conservation law. As can be seen in
Fig.  \ref{FigureThree}, two processes compete with each other:
kinks annihilate mutually $(2X \rightarrow 0)$ and  already existing
kinks branch into an odd number of kinks  ($X \rightarrow 3X,
5X, \ldots$). Both processes resemble a branching annihilating 
walk with an even number of offspring. 
This branching process has a continuous
phase transition that belongs to  the so-called 
parity-conserving (PC) universality class. Phase transitions of this
type have been observed in a variety of models, including certain
probabilistic cellular automata \cite{GrassbergerAB}, nonequilibrium
kinetic Ising models with combined zero- and infinite-temperature
dynamics \cite{Menyhard}, interacting monomer-dimer models \cite{MonDim},
branching-annihilating random walks \cite{BARW} and certain lattice
models with two absorbing states \cite{TwoAbsStates}. In all these models
the symmetry appears either as a parity conservation law or as an explicit
$Z_2$-symmetry among different absorbing phases. A field theory
describing PC transitions is currently developed in \cite{Cardy}.

The PC universality class is characterized by the exponents $\delta=0.285(5)$,
$\eta=0.00(1)$, $z=1.15(1)$, and $\beta=0.92(2)$. In fact, repeating
the numerical simulations described above for $J/k_BT=0.25$ and
$\lambda^*=0.82(1)$ (see Fig. \ref{FigureTwo}), we obtain the estimates
$\delta=0.295(10)$,
$\eta=0.01(2)$, $z=1.17(3)$, and $\beta=0.86(5)$, which are in
fair agreement with the known values. 
We therefore conclude that the DS transition observed for the 
dynamics of eq. (\ref{MixDynamics}) belongs to the PC universality class.
Furthermore, our findings imply that the DS transitions observed 
\cite{GrassJPA} for
the 2-$d$ Ising model with Glauber dynamics should also exhibit PC 
exponents (remember: $d=2$) in zero field, and {\it cross over} to
(2-$d$) DP values when a field is switched on. 

\vspace{3mm}
We thank D. Stauffer for sharing with us his knowledge of the DS
literature and for encouragement. This work was supported by 
The Minerva Foundation and by the Germany-Israel Science Foundation (GIF).
%
%
%


\begin{thebibliography}{99}
\bibitem{Kauffman}
	S.A. Kauffman, J. Theor. Biol. {\bf 22}, 437 (1969).
\bibitem{z}
        P. Grassberger, Physica {\bf A 214}, 547 (1995).
\bibitem{Creutz}
	M. Creutz, Ann. Phys. {\bf 167}, 62 (1986).
\bibitem{Stanley}
	H. Stanley, D. Stauffer, J. Kertesz and H. Herrmann,
	Phys. Rev. Lett. {\bf 59}, 2326 (1987).
\bibitem{DerWeis}
	B. Derrida and G. Weisbuch, Europhys. Lett. {\bf 4}, 657 (1987).
\bibitem{Mariz}
	A. M. Mariz, H. J. Herrmann and L. de Arcangelis, 
	J. Stat. Phys. {\bf 59}, 1043 (1990).
\bibitem{JandArc}
	N. Jan and L. de Arcangelis,
	Ann. Rev. Comp. Phys. {\bf 1},1 (ed. D. Stauffer,
	World Scientific, Singapore 1994).
\bibitem{GrJSP79}
	P. Grassberger, J. Stat. Phys. {\bf 79}, 13 (1995).
\bibitem{PreviousPaper}
	H. Hinrichsen, S. Weitz, and E. Domany, preprint cond-mat/9611085,
	submitted to J. Stat. Phys.
\bibitem{GrassJPA}
	P. Grassberger, J. Phys. A {\bf 28}, L 67 (1995).
\bibitem{Vojta}
	T. Vojta, preprint cond-mat/9610084 (1996).
\bibitem{Costa}
	U. M. S. Costa, J. Phys. A {\bf 20}, L 583 (1987).
\bibitem{LaCaer}
	G. La Caer, J. Phys. A {\bf 22}, L 647 (1989); 
	Physica A {\bf 159}, 329 (1989).
\bibitem{DomanyKinzel}
	E. Domany and W. Kinzel, Phys. Rev. Lett. {\bf 53}, 447 (1984).
\bibitem{DynamicMC}
	P. Grassberger and A. de la Torre, Ann. Phys. (N.Y.) 
		{\bf 122}, 373 (1979).
\bibitem{JensenPRL96}
        I. Jensen, Phys. Rev. Lett. {\bf 77}, 4988 (1996).
\bibitem{GrassbergerAB}
	P. Grassberger, F. Krause, and T. von der Twer,
		J. Phys. {\bf A 17} L105 (1984);
	P. Grassberger, J. Phys. {\bf A 22}, L1103 (1984).
\bibitem{Menyhard}
	N. Menyh\'ard, J. Phys. {\bf A 27}, 6139 (1994);
	B. Menyh\'ard and G. \'Odor, J. Phys. {\bf  A 29}, 7739 (1996).
	The kinetic Ising models introduced in these papers have mixed zero-
	and infinite-temperature dynamics and are different from the usual
	Ising model at finite temperature discussed in the present work.
\bibitem{MonDim}
	H. H. Kim and H. Park,
		Phys. Rev. Lett. {\bf 73}, 2579 (1994);
	H. Park, M. H. Kim, and H. Park, 
		Phys. Rev. {\bf E 52}, 5664 (1995).
\bibitem{BARW}
	I. Jensen, J. Phys. {\bf A 26} 3921 (1993);
	D. ben-Avraham, F. Leyvraz, and S. Redner,
		Phys. Rev. {\bf E 50}, 1843 (1994);
	I. Jensen, Phys. Rev. {\bf E 50} 3623 (1994);
	D. Zhong and D. ben-Avraham, 
		Phys. Lett. {\bf A 209}, 333 (1995).
\bibitem{TwoAbsStates}
	H. Hinrichsen, Phys. Rev. {\bf E 55}, 219 (1997).
\bibitem{Cardy}
	J. Cardy and U. C. T\"auber, Phys. Rev. Lett. {\bf 77}, 4780 (1996).
\end{thebibliography}
\end{document}